\documentclass[useAMS,usenatbib,usedcolumn,usegraphicx]{mn2e}

\def\etal{{\it et\thinspace al.}\ }
\usepackage[dvips]{graphics}
\usepackage[dvipsnames,usenames]{color}

\newcommand{\be}{\begin{equation}}
\newcommand{\ee}{\end{equation}}

\title{[O II] line ratios}
\author[Anil K. Pradhan, Maximiliano Montenegro, Werner Eissner]
       {Anil K. Pradhan$^1$, Maximiliano Montenegro$^{1,2}$, Sultana N.
       Nahar$^1$, Werner Eissner$^3$\\
       $^1$ Department of Astronomy,$^2$ Department of Education,
 The Ohio State University, Columbus, OH 43210, USA,\\$^3$ Institut
 f\"ur Theoretische Physik, Teilinstitut 1, 70550 Stuttgart, Germany}
\date{Accepted  xxxxxx 
      Received xxxxxx;
      in original form xxxxxx}

\pagerange{\pageref{firstpage}--\pageref{lastpage}}
\pubyear{2004}

\def\LaTeX{L\kern-.36em\raise.3ex\hbox{a}\kern-.15em
    T\kern-.1667em\lower.7ex\hbox{E}\kern-.125emX}

\begin{document}

\maketitle

\label{firstpage}

\begin{abstract}
                 Based on new 
calculations we reconfirm the low and high density limits on the
forbidden fine structure line ratio [{\sc O\,ii}] $I(3729)/I(3726)$:
$\lim_{N_{\rm e}\to 0}$ = 1.5 and $\lim_{N_{\rm e}\to \infty}$ = 0.35\@.
Employing [{\sc O\,ii}] collision strengths calculated
using the Breit-Pauli R-matrix method we rule out any significant
deviation due to relativistic effects from these canonical values.
The present results are in
substantial agreement with older calculations by Pradhan (1976)
and validate the extensive observational analysis of gaseous nebulae by
Copetti and Writzel (2002) and Wang \etal (2004) that reach the same
conclusions. The present theoretical results and the recent observational
analyses differ significantly from the calculations by MacLaughlin and
Bell (1998) and Keenan \etal (1999). The new maxwellian averaged
effective collision strengths are presented for the 10 transitions among
the first 5 levels to enable computations of [{\sc O\,ii}] line ratios.
\end{abstract}


\section{INTRODUCTION}
 
  The most prominent density diagnostics in astrophysics is due to the
  forbidden fine structure lines [{\sc O\,ii}] $\lambda\lambda$ 3729,\,3726
  and [{\sc S\,ii}] $\lambda\lambda$ 6716,\,6731. Their utility stems
  from several factors such as:
  (A) they respectively lie at the blue and the red ends of the optical
  spectrum, (B) their atomic structure and hence the density dependence
  is essentially the same, and (C) they are quite strong in the spectra
  of most {\sc H\,ii} regions owing to the relatively large abundances of
  oxygen and sulphur. Seaton and Osterbrock
  (1957) have described the basic physics of these forbidden transitions.
  High-accuracy calculations using the then newly developed computer
  programs based on the close-coupling method IMPACT (Eissner and Seaton
  1972,\,1974, Crees \etal 1978) were later carried out by Pradhan (1976;
  hereafter P76) for the collision strengths, and by Eissner and Zeippen
  (1981) and Zeippen (1982) for the transition probablities. These atomic
  parameters subsequently enabled a consistent derivation of electron 
  densities from observations of [{\sc O\,ii}] and [{\sc S\,ii}] lines in
  a wide variety of {\sc H\,ii} regions (e.\,g. Kingsburgh \& English
  1992, Aller \& Hyung 1995, Aller \etal 1996, Keyes \etal 1990).

   More recently, McLaughlin and Bell (1998; hereafter MB98) repeated the
   [{\sc O\,ii}] calculations of collision strengths using the R-matrix
   method (Burke \etal 1971; Berrington \etal 1995), also based on the
   close coupling approximation and widely employed for a large number
   of atomic calculations ({\em The Opacity Project Team 1995}, Hummer
   \etal 1993). They included a much larger target wavefunction expansion
   than P76, and relativistic effects not considered in the earlier
   calculations. Their electron impact collision strengths and rate
   coefficients were markedly different for the relevant transitions
   $^4$S$^{\rm o}_{3/2} \to ^2$D$^{\rm o}_{5/2,3/2}$ than P76,
   which leads to the theoretical density diagnostic line ratio
   $I(3729)/I(3726)$ to be up to 30\% higher and $\approx$ 2.0 in the
   low-density limit $\lim_{N_{\rm e}}\to 0$. Keenan \etal (1999;
   hereafter K99) recomputed the [{\sc O\,ii}] line ratios to analyze 
   several planetary nebulae using these MB98 results.   

    However, other extensive observational studies (e.\,g. Copetti and
    Writzel 2002, Wang \etal 2004) have noted the discrepancy between
    electron densities derived from [{\sc O\,ii}] and other 
    density indicators, notably the [{\sc S\,ii}] $\lambda\lambda$
    6716,\,6731\@. In particular the recent analysis of a sample of over
    a hundred nebulae by Wang \etal (2004) shows that the collision
    strengths of MB98 are not supported by observations, and
    that the earlier results of P76 are to be preferred.
    But these observational studies leave open the question of what
    precisely are the collision strengths. Given that P76 used a small
    basis set to describe the {\sc O\,ii} target, considered
    no relativistic effects, and could not fully resolve the resonance
    structures in the collision strengths owing to computational
    constraints, it seems puzzling that the new MB98 results which do
    account for all of these factors appear to be inccurate.

     To address this important issue and to resolve the outstanding
     discrepancy, we recently undertook new calcultions for [{\sc O\,ii}]
     using the same Breit-Pauli R-matrix method as employed by
     MB98 and including relativistic effects.  While the details
     of the atomic calculations and comparison of  collision strengths
     with different basis sets of target wavefunctions will be
     presented elsewhere (Montenegro \etal 2005), we present
     the final results of astrophysical interest in this {\it Letter}.

     \section{Theory and computations}

 Forbidden lines are often sensitive to ambient electron density; as
 the Einstein spontaneous decay rates of the upper levels are small,
 they may be collisionally excited to other levels by electron impact
 before radiative decay (Osterbrock 1989, Dopita and Sutherland 2003). 
 This is likely to happen when there is a pair of lines originating
 from closed spaced metastable energy levels, especially in ions of the
 2p$^3$ and the 3p$^3$ outer electronic configurations as exemplified by
 {\sc O\,ii} and {\sc S\,ii}\@. The first five levels are:
 $^4$S$^{\rm o}_{3/2},\,^2$D$^{\rm o}_{5/2,\,3/2},\,^2$P$^{\rm
 o}_{3/2,\,1/2}$\@. The pair of [{\sc O\,ii}] transitions of interest are
 $^2$D$^{\rm o}_{5/2,\,3/2}\longrightarrow ^4$S$^{\rm o}_{3/2}$ at
 $\lambda$\,3729 and $\lambda$\,3726 respectively. 
 
 The basic physics of the limiting values of the line ratio
 $I(3729)/I(3726)$ is quite simple. At low electron densities every
 excitation to the two metastable levels $^2$D$^{\rm o}_{(5/2,\,3/2)}$
 is followed  by a spontaneous decay back to the ground level
 $^4$S$^{\rm o}_{3/2}$
 since the collisional mixing rate among the two excited levels is
 negligible. In that case the line ratio in principle must be equal to the 
 ratio of the excitation rate coefficients
 \be\lim_{N_{\rm e}\to 0} \frac{I(3729)}{I(3726)} =
\frac{q(^4{\rm S}^{\rm o}_{3/2}-^2{\rm D}^{\rm o}_{5/2})}
 {q(^4{\rm S}^{\rm o}_{3/2}-^2{\rm D}^{\rm o}_{3/2})},\ee
where the excitation rate coefficient $q_{ij}$ is
 \be q_{ij}(T) = \frac{8.63 \times 10^{-6}{\rm cm}^3{\rm/s}\cdot
 \exp(-E_{ij}/kT_{\rm e})} {g_i \sqrt{T_{\rm e}/{\rm K}}}
 \Upsilon_{ij}(T_{\rm e}),\ee 
 where g$_i$ is the statistical weight of the initial level and the
 quantity $\Upsilon_{ij}$ is the Maxwellian averaged collision strength:
 \be \Upsilon_{ij}(T) = \int_{E_j}^{\infty} \Omega_{ij}(E)
 \exp(-E/kT_{\rm e})\ {\rm d}(E/kT_{\rm e}). \ee

  If relativistic effects are negligible then
  the collision strengths may be calculated in $LS$ coupling, and an
  algebraic transformation may be employed to obtain the fine structure
  collision strengths. This was the procedure employed in P76.
  The ratio of fine structure $LSJ$ to $LS$ collision collision is
  especially simple when the lower level has either $L$ or $S$ = 0,
  such as for {\sc O\,ii} and {\sc S\,ii}, i.\,e.\

  \be \frac{\Omega(SLJ-S'L'J')}{\Omega(SL-S'L')} =
  \frac{2J'+1}{(2S'+1)(2L'+1)}\,.\ee

  If the excited levels are so closely spaced that the excitation rates
  have virtually the same temperature dependence, the line ratio would then
  be equal to the ratio of the statistical weights (2J'+1) of the upper
  levels $^2$D$^{\rm o}_{5/2,\,3/2}$, i.\,e.\ 6/4. If, however, relativistic
  mixing is significant then the line ratio will depart from the $LS$
  coupling value. That was the contention of MB98 and K99. 

   Therefore we carry out the present calculations including
   relativistic effects and with a suitably large target wavefunction
   expansion. The Breit-Pauli R-matrix (BPRM) calculations with 
   different trial basis sets 
   are described in detail in another paper (Montenegro \etal 2005).
    In the present calculations we employ a 16-level target:
   1s$^2$2s$^2$[2p$^3(^4$S$^{\rm o}_{3/2}, ^2$D$^{\rm o}_{5/2,\,3/2/}, 
   ^2$P$^{\rm o}_{3/2,\,1/2})$; 2s2p$^4(^4$P$_{5/2,\,3/2,\,1/2},
   ^2$D$_{5/2,\,3/2})$, 2p$^2$3s$(^4$P$_{1/2,\,3/2,\,5/2}, 
   ^2$P$_{1/2,\,3/2})$, 2s2p$^4(^4$S$_{1/2})]$. This close coupling
    expansion is more than sufficient to obtain accurate collision
    strengths for the first 5 levels. 
   All resonance structures up to the highest target threshold
   energy $E($2s2p$^4(^4$S$_{1/2})$) = 1.7829 Ry are resolved. The last
   threshold lies sufficiently high to ensure that resonance and
   coupling effects in
   the collision strengths are fully accounted
   for in the excitation of the first 5 levels considered in the
   collisional-radiative model to compute the line ratios. Collision
   strengths at energies 1 Ry higher than the 
   highest of the first 5 levels
   ($E(^2$P$^{\rm o}_{1/2})$ = 0.369 Ry) contribute negligibly to the rate
   coefficients; at $T$ = 20,000\,K the maxwellian factor $\exp(-E/kT)
   \approx$\ e$^{-8}$, and decreasing accordingly for $E>1$\,Ry in Eq. (3). 
   
   In the calculation of [{\sc O\,ii}] line ratios we employ the
   transition probabilities from Zeippen (1982). Eissner and Zeippen
   (1981) computed the $A$-values for [{\sc O\,ii}] transitions taking full
   account of the magnetic dipole M1 operator, and they showed that in the
   high density limit the line ratio
\be\lim_{N_{\rm e}\to\infty} \frac{I(3729)}{I(3726)} = \frac{6}{4}
\frac{A(^2{\rm D}^{\rm o}_{5/2}-^4{\rm S}^{\rm o}_{3/2})}
     {A(^2{\rm D}^{\rm o}_{3/2}-^4{\rm S}^{\rm o}_{3/2})} = 0.35. \ee

\section{Results and discussion}
\begin{figure*}
\begin{minipage}{148mm} 
\centering
\includegraphics[width=148mm,keepaspectratio]{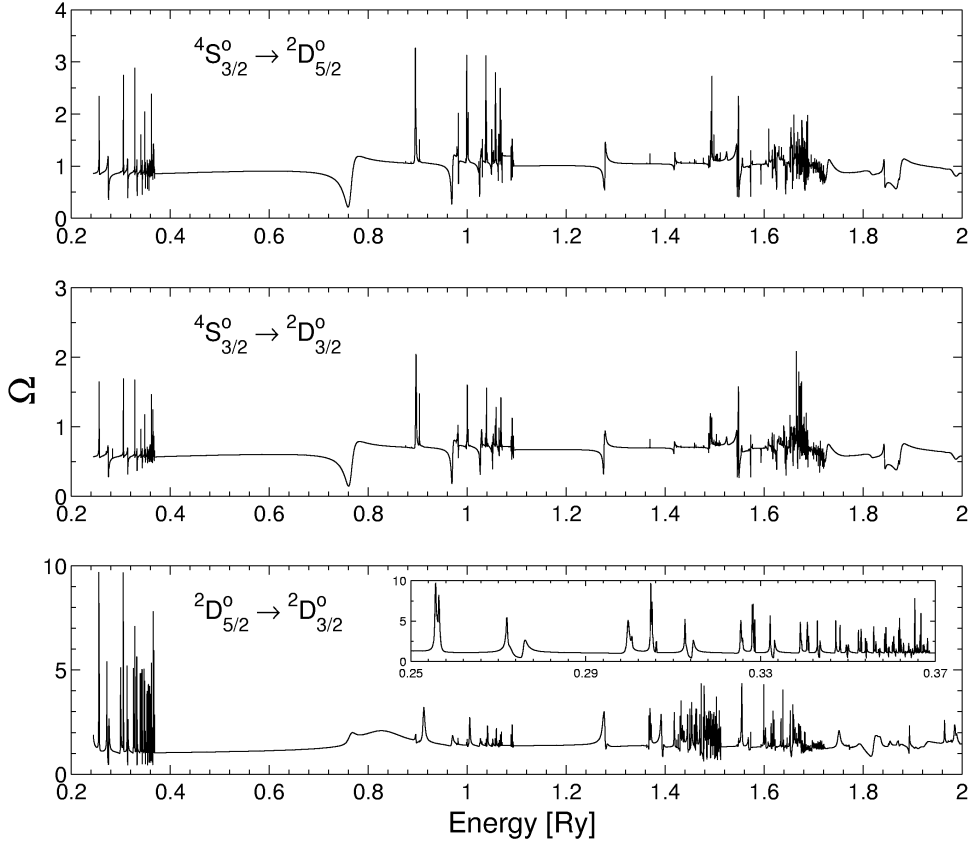}
\caption{Collision strengths for the fine structure transitions
associated with the [{\sc O\,ii}] line ratio 3729\AA/3726\AA\@. Note that
$\Omega(^4$S$^{\rm o}_{3/2}-^2$D$^{\rm o}_{5/2})/
 \Omega(^4$S$^{\rm o}_{3/2}-^2$D$^{\rm o}_{3/2}) = 1.5$ througout.
There is significant resonance enhancement in the collisional mixing
transition $^2$D$^{\rm o}_{5/2}-^2$D$^{\rm o}_{3/2}$; the inset shows the
near-threhold resonances on an expanded scale. \label{coll}}
\end{minipage}
\end{figure*}

 Fig.\ \ref{coll} shows the fine structure BPRM collision strengths
 $\Omega(^4$S$^{\rm o}_{3/2}-^2$D$^{\rm o}_{5/2}),
  \Omega(^4$S$^{\rm o}_{3/2}-^2$D$^{\rm o}_{3/2})$ and
 $\Omega(^2$D$^{\rm o}_{5/2}-^2$D$^{\rm o}_{3/2})$.
 These figures appear to be the first clear presentation of the resonances
 in these collision strengths. P76 did not present detailed resonance
 structures except in the near-threshold region of
 $\Omega(^2$D$^{\rm o}_{5/2}-^2$D$^{\rm o}_{3/2})$.
 MB98 plotted these on an energy scale up to 5\,Ry, well above the
 resonance region up to $\sim$ 2\,Ry, but which does not exhibit the
 resonances in detail to enable comparison. An interesting feature clear
 from Fig.\ \ref{coll} is that the resonances do not play a large role in
 $\Omega(^4$S$^{\rm o}_{3/2}-^2$D$^{\rm o}_{5/2})$ and
 $\Omega(^4$S$^{\rm o}_{3/2}-^2$D$^{\rm o}_{3/2})$ and hence the rate
 coefficients for these transitions. Although they are significant in
 $\Omega(^2$D$^{\rm o}_{5/2}-^2$D$^{\rm o}_{3/2})$, collisional mixing via
 this transition is not important in the low density limit, which therefore
 depends only on the ratio of the transitions from the ground state
 $^4$S$^{\rm o}_{3/2}$ to the $^2$D$^{\rm o}_{5/2,\,3/2}$ levels. We find
 that this ratio is constant at 6/4 throughout the energy range under
 consideration, resonant or non-resonant values. Therefore we do not find
 any signifant evidence of relativistic effects, which would manifest
 itself in a departure from this ratio. Remarkably the present total sum
 $\sum_{J=5/2,3/2}\Omega(^4$S$^{\rm o}_{3/2}-^2$D$^{\rm o}_J)=1.42$,
 compared to the $LS$ coupling P76 value of 1.31, and an even earlier
 value of 1.36 obtained by Saraph, Seaton, and Shemming (1969).

  Table 1 give the Maxwellian averaged collision strengths $\Upsilon(T)$
  for the 5-level [{\sc O\,ii}] model. At 10,000\,K we obtain
  $\Upsilon(^4$S$^{\rm o}_{3/2}-^2$D$^{\rm o}_{3/2}) = 0.585$, in good
  agreement with the earlier P76 value of 0.534, about 9\,\% lower, but 
  considerably higher than the M98 value of 0.422 (quoted in K99)
  which is 28\% lower than the new value. More importantly our results
  disagree with MB98 for the ratio discussed above.
   It is this ratio that is responsible for the K99 line ratio
   $I(3729)/I(3726)$ to be about 30\% higher ($\sim2.0$) than the expected
   low-density limit of 1.5, as shown in Fig.\ 2.

%
%
\begin{table*}
\begin{minipage}{148mm}
\caption{Effective Maxwellian averaged collision strengths}
\begin{tabular}{l@{$\,\to\,$}l......}
\hline
\multicolumn{2}{c}{ Transition} & \multicolumn{1}{c}{$\Upsilon$(1000\,K)}
 & \multicolumn{1}{c}{$\Upsilon(5000\,\rmn{K})$}
 & \multicolumn{1}{c}{$\Upsilon(10000\,\rmn{K})$}
 & \multicolumn{1}{c}{$\Upsilon(15000\,\rmn{K})$}
 & \multicolumn{1}{c}{$\Upsilon(20000\,\rmn{K})$}
 & \multicolumn{1}{c}{$\Upsilon(25000\,\rmn{K})$}\\

\hline
$^4$S$^\rmn{o}_{3/2}$ & $^2$D$^\rmn{o}_{5/2}$ &
 0.864 &   0.885 & 0.883 &   0.884 &  0.885 & 0.888 \\

$^4$S$^\rmn{o}_{3/2}  $ & $ ^2$D$^\rmn{o}_{3/2}$&
 0.590 & 0.587 & 0.585 & 0.585 & 0.585 & 0.588 \\

$^2$D$^\rmn{o}_{5/2}$ & $^2$D$^\rmn{o}_{3/2}$ &
1.618 & 1.518 &  1.426 & 1.365 & 1.324 & 1.320\\

$^4$S$^\rmn{o}_{3/2} $ & $  ^2$P$^\rmn{o}_{3/2}$ &
 0.299 & 0.307 &  0.313 & 0.318 & 0.322 & 0.327\\

$^2$D$^\rmn{o}_{5/2} $ & $  ^2$P$^\rmn{o}_{3/2}$ &
 0.912 & 0.928 &   0.946 & 0.971 & 1.000 & 1.030\\

$^2$D$^\rmn{o}_{3/2} $ & $  ^2$P$^\rmn{o}_{3/2}$ &
0.571 & 0.589 &   0.605 & 0.624 & 0.644 & 0.664 \\ 

$^4$S$^\rmn{o}_{3/2} $ & $  ^2$P$^\rmn{o}_{1/2}$ &
 0.148 & 0.151 &  0.152 & 0.154 & 0.156 & 0.158 \\

$^2$D$^\rmn{o}_{5/2} $ & $  ^2$P$^\rmn{o}_{1/2}$ &
 0.383  & 0.392 &  0.402  & 0.414   & 0.428   & 0.441\\

$^2$D$^\rmn{o}_{3/2} $ & $   ^2$P$^\rmn{o}_{1/2}$ &
 0.376 & 0.386 &  0.397 & 0.409 & 0.423 & 0.437 \\ 

$^2$P$^\rmn{o}_{3/2} $ & $  ^2$P$^\rmn{o}_{1/2}$ &
0.277& 0.284 &  0.291 & 0.300 & 0.310 & 0.321\\
\hline
\end{tabular}
\end{minipage}
\end{table*}

 Comparing the present relativistic BPRM results for effective collision 
 strengths with the LS coupling results of P76 we find good agreeement,
 mostly within a few percent, with the notable exception of
 $\Upsilon(^2$D$^{\rm o}_{5/2}-^2$D$^{\rm o}_{3/2})$.
 Owing to the more extensive delineation of resonance structures in the
 present calculations (Fig.\ 1), the $\Upsilon$ value is much higher. For 
 example, at $T$ = 10,000\,K the P76 value is 1.168 compared to the
 present value of 1.426 in Table 1. On the other hand the present
 $\Upsilon(^2$P$^{\rm o}_{3/2}-^2$P$^{\rm o}_{1/2})$ = 0.291 agrees well
 with the P76 value 0.287 at $T$ = 10,000\,K.

 Fig.\ 2 shows that the present collision strengths yield the line ratio
 $R_1 = I(3729)/I(3726)$, which approaches the low- and high-density
 limits exactly. The difference is not discernible when we use the P76
 values. On the other hand the difference with MB98 is quite pronounced
 and approaches $\sim$30\% in the low-density limit. The temperature 
 variation between $T$ = 10,000\,K (solid line) and $T$ = 20,000\,K (dashed
 line) is also small, demonstrating the efficacy of this ratio as an
 excellent density diagnostic.

\begin{figure} 
\centering  
\includegraphics[width=\columnwidth,keepaspectratio]{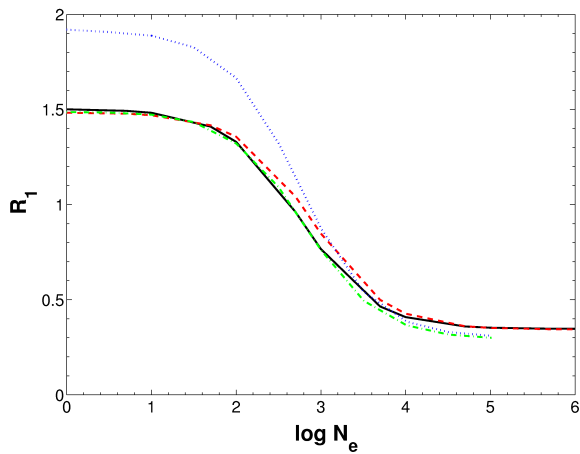}
\caption{[{\sc O\,ii}] line ratio $I(3729)/I(3726)$ vs electron density
N$_{\rm e}$: present results --- solid line,
 Pradhan (1976) -.- dot-dashed line
(nearly indistinguishable from the solid line),
McLaughlin and Bell (1998) .... dotted line, at $T$ = 10,000\,K. The
dashed line is the line ratio at $T$ = 20,000\,K. \label{lineratio}}
\end{figure}

\section{Conclusion}
 We have carried out new relativistic Breit-Pauli R-matrix calculations
 for the [{\sc O\,ii}] transitions responsible for the important density
 diagnostic line ratio 3729\AA/3726\AA\@. We find no evidence of any
 significant departure from the earlier $LS$ coupling results of Pradhan
 (1976). The line ratios derived from the present results also agree with
 the canonical limits expected on physical grounds. The new results are in
 considerable disagreement with the calculations of McLaughlin and Bell
 (1989) and the line ratios of Keenan \etal (1999). We also reconfirm
 the observational analyses of Copetti and Writzel (2002) and Wang \etal
 (2004).

\section*{Acknowledgments}
 AKP would like to thank Prof.\ Don Osterbrock for first pointing out
 this problem. The computational work was 
carried out on the Cray X1 at the Ohio Supercomputer Center in Columbus 
Ohio. This work was partially supported by a grant from the U.S.
National Science Foundation.

\label{lastpage}

\end{document}